# A new PCA-based utility measure for synthetic data evaluation

Fida K. Dankar[1*]; Mahmoud K. Ibrahim[2]
[1]New York university Abu Dhabi, UAE
[2]KU Leuven, Belgium
*corresponding author: fida.dankar@uaeu.ac.ae

**Abstract.** Data synthesis is a privacy enhancing technology aiming to produce realistic and timely data when real data is hard to obtain. Utility of synthetic data generators (SDGs) has been investigated through different utility metrics. These metrics have been found to generate conflicting conclusions making direct comparison of SDGs surprisingly difficult. Moreover, prior research found no correlation between popular metrics, concluding they tackle different utility-dimensions. This paper aggregates four popular utility metrics (representing different utility dimensions) into one using principal-component-analysis and checks whether the new measure can generate synthetic data that perform well in real-life. The new measure is used to compare four well-recognized SDGs.

## 1 Introduction

### 1.1 Background

Healthcare providers, governments and private organizations are collecting large amounts of patient data in electronic records. These new generation of datasets are voluminous and include complex data from multiple sources entailing both phenotype (medical records, clinical studies, etc.) and genotype (variation screening at first, now increasingly shifting to Whole Exome and Whole Genome sequencing) [1]. While such datasets have the potential to make breakthroughs in the health field if explored by scientists, they are generally not accessible for the broad research community due to ethical challenges such as privacy concerns. Even when access to good quality individual-level data is possible, it entails a lengthy application process with strict legal requirements to ensure proper usage and proper protection of the data [2]. A recent report from the US Government Accountability Office identifies data privacy as a major barrier to the application of artificial intelligence and machine learning in healthcare [3]. The report reveals that a lot of effort goes into accessing and curating data to make it usable for machine learning applications, thus severely delaying the pace of research and creating a barrier to data profit and progress in health care.

Synthetic data generation is rapidly emerging as a promising privacy enhancing technology for secondary use of health data [4]–[8]. Synthetic data is artificial data that is simulated from real data to mimic its statistical properties. A model that captures the patterns in the real data and its statistical properties is built. Then the model is used to generate synthetic data. Synthetic data is gaining popularity in the health field as it can solve the issues of broad data availability as well as delays in acquiring data. In a recent WSJ article, Gartner predicts that, 60% of the data used for artificial intelligence and analytics will be synthetically generated by 2024 [9].

Measuring whether synthetic data retains the information required for analysis is far from trivial and is an active area of research. It is commonly understood that for inferences derived from synthetic data to be valid, the models used for generating them should conform to the models that generated the real data. The assessment of this condition is often done by measuring the utility of the released data. Utility attempts to quantify the difference between the original and masked data by checking whether certain properties of the original dataset are preserved in the masked one [10]. Dreschler and Reiter classify available utility metrics into two categories, narrow and broad [11]. Narrow metrics (also referred to as analysis-specific metrics) assess the ability of the released data in replicating a specific analysis done on the real data (for example coefficients of a machine learning model), as such they may not provide desired results for other kinds of analysis. Narrow metrics can be useful in evaluating utility when the type of analysis to be applied is known ahead of time. Broad measures (also known as general measures or metrics) try to capture features of the entire dataset by quantifying some kind of statistical distance between the original and the masked datasets.

Multiple broad metrics are available in the literature, with each focusing on the preservation of a particular statistical measure. These metrics have been found to generate conflicting conclusions making the assessment of synthetic data utility difficult. It is a common practice to use propensity score to estimate utility [12], propensity measures the distributional similarity between the original and released dataset and is advocated as the best measure for synthetic data evaluation [10], [13].

### 1.2 Contributions

Prior research categorized the available broad measures into 3 utility dimensions and proved that the metrics across these dimensions are not correlated [14], which implies that all metrics are required when evaluating the overall utility of synthetic data (SD). Building on these ideas, we introduce a new utility measure that unifies multiple broad measures belonging to the different utility dimensions using principal component analysis (PCA). The new measure is an attempt to represent all dimensions of utility in one measure. The paper then evaluates whether the new measure can be used to generate synthetic datasets that perform well in real life in comparison with the commonly used propensity. Both measures are then used to rank four well-recognised (open source) synthetic data generators (SDG).

The paper proceeds as follows: Section 2 provides an overview of broad utility measures, Section 3 presents the different SDGs used in the study, introduces the novel PCA based utility measure and presents the materials and methods used for synthetic data generation and evaluation, Section 4 presents the results of the experiments and Section 5 concludes the paper.

## 2 Data utility

In [14], the authors classify broad utility metrics into 3 main categories based on the statistics they attempt to preserve: Attribute fidelity, bivariate fidelity and population fidelity metrics (see Figure 1). The three categories are described below:

**Attribute fidelity metrics** (or univariate fidelity) focus on the basic structural similarity of the dataset. They require each attribute in the masked data have similar structure and similar fundamental aggregated statistics (same variable types, formats, names, means, and ranges) as well as similar univariate distributions for both, the continuous and discrete variables (i.e. similarity in the marginal distributions of the original and released datasets). Such measures are commonly used in the synthesis literature and are necessary to determine whether the same code can be applied to both datasets without producing syntactical errors [15], [16]. Measures commonly used are Hellinger distance [17] and Kullback-Leibler divergence [18]. Univariate fidelity is considered the minimum requirement for SD to be useful, implying that all marginal distributions of original and synthetic datasets should be matching [15].

**Bivariate fidelity metrics** focus on correlations among pairs of variables in the dataset so as to capture the statistical dependency structure of the original and released datasets. Retaining such structure in the synthetic dataset ensures that the underlying relationships between the attributes are preserved (for example symptoms are attributed to the right diagnosis and employment status is attributed to the appropriate age) such association is crucial to ensure truthful representation of the original dataset. Bivariate fidelity can be measured using pairwise correlation plots such as heat maps [19], [20].

**Population fidelity metrics** (or multivariate fidelity) are the most popular metrics for the overall evaluation of masked data. They reflect large-scale features of the entire distributions of the masked dataset in comparison with the original dataset. These measures are believed to be helpful in allowing a global assessment of how well the final inference might agree with what would have been obtained had the user had access to the original data [10]. Multiple metrics exist in this category, *The log-cluster metric* measures the similarity of the underlying dependency structure in terms of clustering [8], [12]. Real and synthetic data are merged into one dataset, a cluster analysis with a fixed number of clusters is performed on the merged dataset, placing records into clusters of similar values. Finally, a metric is calculated to reflect the distribution of the synthetic dataset across the different clusters. If the allocation to the different clusters is similar for synthetic and real, then this suggests similar distributions. *Difference in Empirical distributions type metrics* measure the differences between the empirical cumulative distribution functions calculated for real and synthetic datasets. Kolmogorov-Smirnov type statistics [12] estimate the empirical distributions difference. They first calculate the discrete empirical distributions of both datasets (from the supplied sample), then they calculate the average square differences between the two. *Distinguishability type metrics* characterize the extent to which it is possible to distinguish the original dataset from the synthesized one. The most prominent distinguishability measure is the propensity score [16], [21]. It involves building a classification model to distinguish between the real and released datasets records. A high utility implies the inability of the model to perform the distinction.

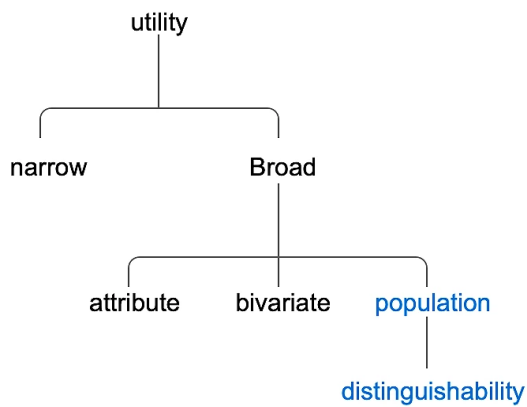

Figure 1. Classification of utility measures

Prior investigation into a correlation between the different utility categories found it to be insignificant [14], the result implies that no category can substitute another, and that all categories are needed to provide an overall assessment of synthetic data utility. Here, we propose a new utility measure, $u_{pca}$, for synthetic data evaluation that embodies measures from each of the aforementioned utility categories and then we investigate how well the new measure correlates with the performance of synthetic data when used to generate machine learning models.

## 3  Methods

Synthetic datasets are generated from a model that is fit to real data using either statistical or machine learning methods [22]. Statistical methods use a family of predefined distributions to fit the real dataset (such as Gaussian models or Bayesian Networks), while machine learning methods build a machine learning model to fit the real dataset (such as classification and regression trees or generative adversarial networks). The generation process is stochastic, which implies that a different synthetic dataset is generated each time the model is used.

We included four publicly available methods for synthetic data generation that are among the most influential work in this area [7]. They include two statistical methods (i) a Bayesian network based data synthesis technique-*DataSynthesizer* [23], and (ii) a copula-based data synthesis technique-*synthetic data vault* [24]. As well as two machine learning methods (iii) a parametric data synthesis technique-*Synthpop parametric* [25], and (iv) a non-parametric tree-based data synthesis technique-*synthpop non-parametric* [20].

The synthetic data vault models the dependency structure between the variables using Copulas [26]. Copula-based models were shown to be effective for generating high quality synthetic data, are easy to use and are computationally efficient when compared to other statistical generators [27]–[29]. DataSynthesizer is another well-cited SDG, it provides a publicly available implementation of a Bayesian network SDG. It has been used in NIST's 2018 Synthetic Data Challenge. It is based on a method published by Zhang et al. [30], but incorporates a number of modifications to improve its data utility and efficiency. Synthpop (parametric and non-parametric) are among the most recognized machine learning generators [13], [31]. They are used extensively for the synthesis of health and social science data [13], [32]–[35]. They are preferred over other deep learning SDGs for smaller datasets like the ones we use in this paper [13], [27]. Other approaches for synthetic data generation exist, some are developed for specific problems or specific data types [36], [37], whereas these techniques are considered more general [7].

The four methods are described in details in the following sub-section.

## 3.1 Synthetic data generators

The first generator, referred to as DataSynthesizer or DS was developed in Python in 2017. It captures the underlying correlation structure between the different attributes by constructing a Bayesian network [23]. The method uses a greedy approach to construct the Bayesian network modeling the correlated attributes. Samples are then drawn from this model to produce the synthetic datasets. DS supports numerical, datetime, categorical and *non-categorical string* data types. Non-categorical strings status is given to categorical attributes with big domains (such as names), they are defined by their length (a range of minimal to maximal length). Non-categorical strings allow DS to generate random strings within the length range during the data generation stage. This feature enables the DataSynthesizer to create datasets that feel like the real sample by including synthetic string data such as names and IDs. DS also supports missing values by taking into consideration the rate of missing values within the dataset. For more details readers are referred to [23], [30].

The second synthesizer, the Synthetic Data Vault (SDV), was developed in 2016 by Patki et al [24]. SDV estimates the joint distribution of the population using a latent Gaussian copula [26]. Given a private dataset $D$, with $d$ attributes, drawn from a population $P$, SDV models the population's cumulative distribution function $F$ from the sample by specifying separately the marginal distributions of the different attributes, $F_1, \ldots, F_d$, as well as the dependency structure among these attributes through the Gaussian Copula. Marginal distributions are inferred from the sample $D$, and Gaussian copula is expressed as a function of the covariance matrix. For more information, the reader is referred to [24]. SDV is implemented in Python, it assumes that all values are numerical. When the assumption does not apply, SDV applies basic pre-processing on the dataset to transform all categorical attributes into numerical values between [0,1] (the value reflects their frequency in the dataset). Datetime values are transformed into numerical values by counting the number of seconds from a given timestamp. And missing values are considered important information and are modeled as null values.

The third and fourth generators come from Synthpop (SP) and were developed in 2015 using the *R* language but are continuously updated with new features [20]. SP generates the synthetic dataset sequentially one attribute at a time by estimating conditional distributions. If $X_i$ denotes the random variable for column $i$, then SP proceeds sequentially as follows: For random variable $X_1$, the marginal distribution, $\Pr(X_1 = x_1)$, is estimated from the first column of $D$. The distribution is then used to generated the entries for the first synthetic data column $(x_1^1, \ldots, x_1^{n_D})^T$. Moving to the second attribute, the conditional distribution $\Pr(X_2|X_1)$ is estimated and used along with the synthesized values of the first column $(x_1^1, \ldots, x_1^{n_D})^T$, to generate the synthetic values: $\Pr(X_2 = x_2|X_1 = x_1)$. The same method repeats until all columns are synthesized. SP presents two methods for the generation of the conditional distributions:
$\Pr\big((X_j = x_j)|(X_1 = x_1, \ldots, X_{j-1} = x_{j-1})\big)$, the default method uses the nonparametric CART algorithm (Classification and Regression Trees), referred to as *SP-np*, and the second parametric method uses logistic regression to generate the conditional distribution, we refer to this case as SP-parametric or *SP-p*. Both parametric and nonparametric methods are included in the comparison. Note that for SP, the order in which the attribute are synthesised affects the utility of the generated synthetic data [15]. The SP algorithm orders attributes according to the number of distinct values in each (attributes with fewer distinct values are synthesized first).

## 3.2 Utility Metrics

We introduce a new utility metric $u_{PCA}$ and evaluate it against propensity score. Propensity score ($p$) is a distinguishability type metric from the population fidelity category, it represents the probabilities of record memberships (original or synthetic). It is commonly used to for synthetic data evaluation and is advocated as the best such measure [10], [20], [38]. It is derived from a metric proposed by Woo et al [12], and altered specifically for synthetic data by Snoke et al [21]. It is cited as the most promising (overall) utility measure [12], [21], the most practical measure [15], a measure that can be used to optimize synthetic generation [13], a valuable measure for comparing different synthesis approaches [39], and a promising measure for general synthetic data utility [12]. Both metrics are described in details in the following subsections.

### 3.2.1 Propensity

To calculate $p$, the original and synthetic datasets are joined in one dataset and a binary indicator is added to each record depending on whether it is real or synthesized. A binary classification model is then constructed to distinguish between real and synthetic records. The model is then used to compute the propensity score $\hat{p}_i$ for each record $i$ (predicted value for the indicator). The propensity score is calculated from the predicted value as follows:

$$p = \frac{1}{N}\sum_i (\hat{p}_i - 0.5)^2$$

Where $N$ is the size of the joint dataset.

The propensity score varies between 0 and 0.25, with 0 indicating no distinguishability between the two datasets. This can happen if the generator overfits the original dataset and creates a synthetic that is indistinguishable from the original one (leading to a score of $\hat{p}_i$ =0.5 for every record). On the other extreme, if the two datasets are completely distinguishable, the propensity score for each record will match the binary indicator, leading to an overall score of 0.25.

### 3.2.2 PCA-based Utility

The new utility metric, $u_{PCA}$, is based on four utility metrics spanning the three broad utility categories (attribute, bivariate and population fidelity). The choice of the metrics from within the categories is based on their acceptance in prior SD literature.

1. We use the **Hellinger distance ($H$)** to assess attribute fidelity. $H$ is commonly used to compare the univariate distributions of the real and synthetic datasets [13], [40]. It has been shown to behave consistently when used to compare real and masked datasets in the context of disclosure control evaluation [41]. Moreover, it is bounded between 0 and 1 which makes it easy to interpret.
   H calculates the univariate distance between corresponding pairs of attributes and returns the average distance across all variables. For each attribute $v_o$ in the real dataset and its corresponding $v_s$ from the synthetic dataset, H is calculated as follows:

   $$H(v_o, v_s) = \frac{1}{\sqrt{2}}\sqrt{\sum_i (\sqrt{p_i} - \sqrt{q_i})^2}$$ for discrete attributes and

   $$H(v_o, v_s) = \frac{1}{\sqrt{2}}\sqrt{\int (\sqrt{f(x)} - \sqrt{g(x)})^2 \, dx}$$ for continuous attributes.

   $p_i$ and $q_i$ correspond to the probability for every distinct value in the original and synthesized datasets, and $f$ and $g$ denote the probability density functions for the original and synthesized columns respectively. The overall Hellinger distance is calculated as the mean Hellinger distance across all variables in each iteration. The smaller the $H$ value, the closer the synthetic dataset is to the real dataset in terms of univariate distributions across all variables.

2. Bivariate fidelity is incorporated through the **pairwise correlation difference ($PCD$)** metric. Bivariate fidelity is often measured through heat maps and correlation plots [7], [19], [23], [42], however, we chose $PCD$ as it provides a numerical assessment of the correlation that can be incorporated within our metric. $PCD$ is intended to measure how much correlation is preserved among pairs of variables by calculating the difference between correlation matrices for real and synthetic data. Smaller values for $PCD$ imply that the synthetic and real data are close in terms of linear correlations across variables. $PCD$ is defined as:

$$PCD(R,S) = ||Corr(R) - Corr(S)||_F\,^1$$

Where $Corr(R)$, and $Corr(S)$ are the correlations matrices for the real and synthetic data respectively. $PCD$ values are lower when both datasets are close in terms of linear correlations across variables.

3. Population fidelity studies the correlation structure between all variables. Such structure is crucial for a reliable representation of the original dataset. As such, we incorporate two metrics from the population fidelity category: propensity and log-cluster. **Propensity** is included as it is the most recognized broad metric. As for **Log-cluster metric ($LC$)**, it is a measure of the similarity of the underlying dependency (causal) structure between the original and synthesized datasets [12]. LC is a well-recognized broad utility measure for synthetic datasets [8], [12]. The real and synthetic datasets are merged and clustering algorithms are applied on the data to partition the observations into *c*lusters, the proportion of real vs synthetic data is then assessed within each cluster. When $LC$ is large, that implies that there are disparities in the proportions of real and synthetic data in the different clusters, implying a difference in the underlying distributions. To calculate LC, the real and synthetic datasets are merged into one dataset, then the k-means combined with k-modes clustering algorithms are applied on the data to partition the observations into *k* clusters. The log-cluster metric is then calculated as follows:

$$LC(R,S) = \log\left(\frac{1}{k}\sum_{i=1}^{k}\left[\frac{n_i^R}{n_i} - \frac{1}{2}\right]^2\right)$$

Where $n_i^R$ is the number of real records within cluster $i$ and $n_i$ is the total number of records in the cluster. When $LC(R,S)$ is large, that implies that there are disparities in the proportions of real and synthetic data in the different clusters, implying a difference in the underlying distributions. The value for *k* is chosen to be $\frac{N}{10}$ (where $N$ is the size of the joint dataset).

The utility metric, $u_{pca}$, takes all four defined metrics and uses PCA to reduce the dimension to 1. PCA is commonly used for dimensionality reduction. It works by projecting each data point onto the first few principal components to obtain lower-dimensional data while preserving as much of the data's variation as possible [43]. In our case, PCA takes four-dimensional points as input and reduces their dimensionality to one, (i.e. to the first principal component). The first component represents a direction that maximizes the variance of the projected data. The process in explained in details later in this section.

### 3.3 Performance measures

Real-life performance of synthetic data is hard to capture as data can be used for a variety of applications. Prior research evaluated the performance of synthetic data using specific applications, particularly supervised machine learning [7], [25], [44], [45]. In supervised machine learning applications, if inferences from the constructed models agree between synthetic and real data, then the synthetic data is said to have high utility. We use a combination of four simple and complex machine learning classification algorithms to assess SD performance. Two parametric algorithms: Logistic regression (LR) and support vector machines (SM), as well as two non-parametric algorithms: random forest (RF) and decision trees (DT).

---

[1] The correlation of 2 columns is assumed to be 0 when one of the columns is single valued. That occurred few times in the generated synthetic datasets when the original real columns were highly imbalanced.

For each synthetic dataset, the 4 different classification models (CM) were trained. Subsequently, the models generated are tested on the testing set of the corresponding real data. Testing on real data allows us to determine how well a model trained on synthetic data will perform on real cases.

### 3.4 Synthetic Data generation Process & Utility Calculation

#### 3.4.1 Synthetic data generation

As mentioned before, synthetic data generation is non-deterministic, so multiple instances generated from the same model will exhibit utility variations. To lower the variability effect, it is a common practice to generate multiple datasets from the same model and use them as one ensemble. Utility of the ensemble is then calculated as the average across the units. In our experiments, we use ensembles of size 5.

Unprocessed real data is randomly divided into 70% training and 30 % testing, synthetic data ensembles are generated from the training datasets and utility metrics are then calculated for each generated synthetic dataset Ensemble. The synthesizing process is depicted in Figure 2 and explained formally in details after presenting the different datasets used in the experiments.

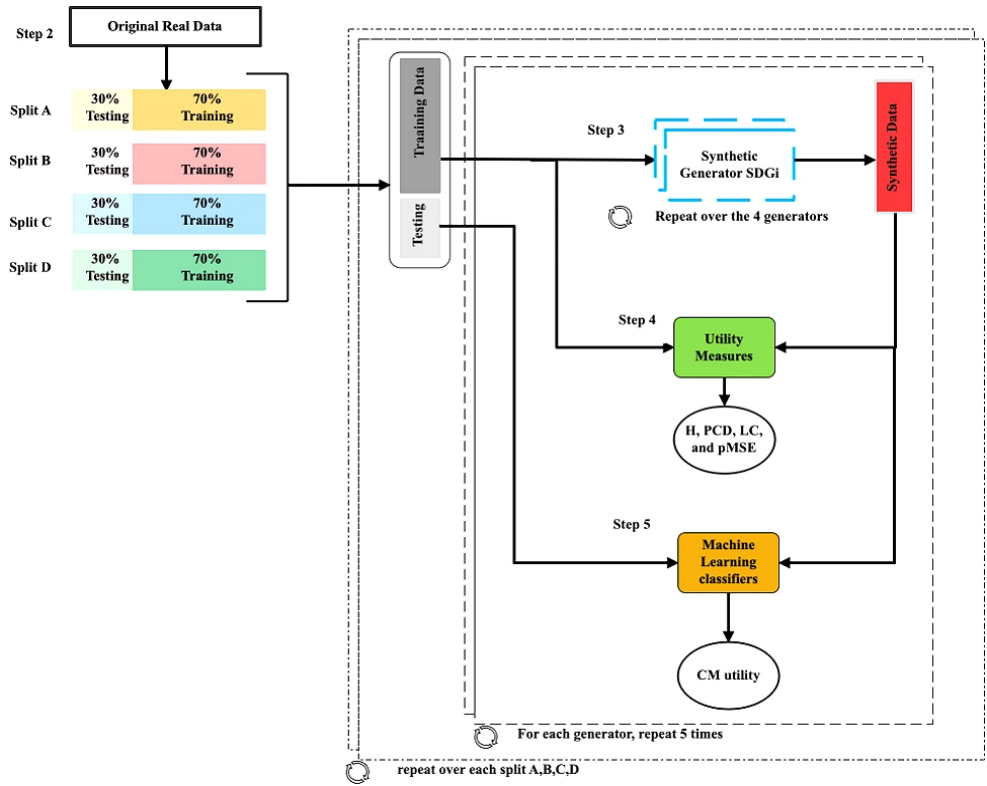

Figure 2. The figure describes the data synthesis and utility calculation processes. The final utility for each [generator, dataset] pairs is the average over the corresponding synthetic data utilities.

The datasets used in the experiments are taken from different sources and are of different size and feature counts. The sources are: University of California Irvine repository [46], OpenML platform [47], Datasphere (DS) [48], Cerner clinical database (CCD) [49] and Kaggle community platform [50]. Details about the datasets are provided in Table 1.

| Dataset name | Acronym | Observations | Number of predictors | Categorical predictors | Number of labels | Synthetic datasets generated | Origin |
|---|---|---|---|---|---|---|---|
| BankNote | $D_1$ | 1,372 | 4 | 0 | 2 | 80 | UCI |
| Titanic | $D_2$ | 891 | 7 | 7 | 2 | 80 | Kaggle |
| Ecoli | $D_3$ | 336 | 7 | 0 | 8 | 80 | UCI |
| Diabetes | $D_4$ | 768 | 9 | 2 | 2 | 80 | UCI |
| Cleveland heart | $D_5$ | 297 | 13 | 8 | 2 | 80 | UCI |
| Adult | $D_6$ | 48,843 | 14 | 8 | 2 | 80 | UCI |
| Breast cancer | $D_7$ | 570 | 30 | 0 | 2 | 80 | UCI |
| Dermatology | $D_8$ | 366 | 34 | 33 | 6 | 80 | UCI |
| SPECTF Heart | $D_9$ | 267 | 44 | 0 | 2 | 80 | UCI |
| Z-Alizadeh Sani | $D_{10}$ | 303 | 55 | 34 | 2 | 80 | UCI |
| Diabetic Data | $D_{11}$ | 101766 | 50 | 36 | 3 | 80 | CCD |
| Colposcopies | $D_{12}$ | 287 | 68 | 6 | 2 | 80 | UCI |
| ANALCATDATA | $D_{13}$ | 841 | 71 | 3 | 2 | 80 | OpenML |
| Mice Protein | $D_{14}$ | 1,080 | 80 | 3 | 8 | 80 | UCI |
| Diabetic Mellitus | $D_{15}$ | 281 | 97 | 92 | 2 | 80 | OpenML |
| Tecator | $D_{16}$ | 240 | 124 | 0 | 2 | 80 | OpenML |
| Colorectal | $D_{17}$ | 690 | 176 | 136 | 2 | 80 | DS NCT00384176 |
| Arrhythmia | $D_{18}$ | 452 | 279 | 116 | 2 | $70^2$ | UCI |
| Scene | $D_{19}$ | 2407 | 293 | 6 | 2 | $60^3$ | OpenML |

**Table 1.** Datasets description

Referring to the 4 synthetic generators (DS, SDV, SP-np, SP-p) as: $SDG_1$, $SDG_2$, $SDG_3$, $SDG_4$ respectively, the experiments are formally defined as follows:

(i) Each real dataset, $D_i$, is randomly split 4 times into 70% training and 30 % testing, where $DT_i^1, \dots, DT_i^4$ denote the training sets and $Dt_i^1, \dots, Dt_i^4$ their corresponding testing sets.

(ii) For each training dataset and synthesizer pairs: [ $DT_i^r, SDG_k$ ], we generate 5 synthetic datasets referred to as datapoints. The five datasets are treated as one unit (or ensemble). A total of 16 ensembles are generated for each dataset $D_i$.

(iii) The 16 ensembles generated are referred to as: $SD_1^i, \dots, SD_{16}^i$, where:

   a. $SD_1^i, \dots, SD_4^i$ correspond to $SDG_1$, $SD_5^i, \dots, SD_8^i$ correspond to $SDG_2$, $SD_9^i, \dots, SD_{12}^i$ correspond to $SDG_3$, and $SD_{13}^i, \dots, SD_{16}^i$ correspond to $SDG_4$.

   b. $SD_1^i, SD_5^i, SD_9^i, SD_{13}^i$ correspond to $DT_i^1$, $SD_2^i, SD_6^i, SD_{10}^i, SD_{14}^i$ correspond to $DT_i^2$, $SD_3^i, SD_7^i, SD_{11}^i, SD_{15}^i$ correspond to $DT_i^3$, and $SD_4^i, SD_8^i, SD_{12}^i, SD_{16}^i$ correspond to $DT_i^3$

   c. Thus $SD_j^i$ corresponds to $SDG_{j div 4+1}$ and $DT_i^{j mod 4}$

(iv) The utility measures $H, PCD, LC$, and $p$, are calculated for each datapoint, and the utility measures for ensemble $SD_j^i$ are calculated as the average over the utility measures of the corresponding datapoints and are denoted by $H_j^i, LC_j^i, p_j^i, PCD_j^i$,

(v) For each real dataset $D_i$, the four CM are trained using the training dataset, and the prediction accuracy is calculated using the corresponding real testing dataset and are denoted by $PA_{LR}^i, PA_{RT}^i, PA_{SM}^i, PA_{RF}^i$. We perform binary classification and multiclass classification, depending on the number of labels of each dataset (see Table 1).

---

[2] generated only 10 datasets using DS due to time efficiency (more than 15 hours was taken to generate each dataset).
[3] generated all synthetic data except for DS due to time efficiency.

(vi) For each datapoint, the prediction accuracy is calculated for the four CM using the corresponding real testing dataset: $Dt_j^r$. The accuracies for the ensemble $(SD)_j^i$ are calculated as the average across the 5 datapoints and are denoted by $PA_{LR,j}^i, PA_{RT,j}^i, PA_{SM,j}^i, PA_{RF,j}^i$ [4]

At the end of the experiments, each synthetic data ensemble (or synthetic data henceforth) $SD_j^i$ will be associated with four metric values: $(H_j^i, LC_j^i, p_j^i, PCD_j^i)$ and four accuracy values: $(PA_{LR,j}^i, PA_{RT,j}^i, PA_{SM,j}^i, PA_{RF,j}^i)$.

### 3.4.2 Utility metrics Calculation

We use two utility metrics to evaluate synthetic data, Propensity score ($p$) and PCA based $u_{pca}$. Here, PCA takes four-dimensional points as input (($H_j^i, LC_j^i, p_j^i, PCD_j^i$)) and reduces their dimensionality to one, i.e. to the first principal component, $u_{pca}{}_j^i$. Ten real datasets are specified for training, implying that all synthetic data generated from these datasets ($10 \times 16$) are used to generate the PCA model. The remaining 9 real datasets are specified for testing. They are used to compare both measures on their ability to predict the performance of synthetic data when used to generate machine learning models. In what follows, $p$ and $u_{pca}$ will be referred to as quality metrics to distinguish them from other metrics.

### 3.5 Set-up

We used AWS [51] Virtual machine, instance type: r5a.8xlarge, having 32 vCPUs with 256 GiB memory for data generation. Multiple packages were used throughout our experiments. For calculating the propensity scores, the ps function from the R twang package was used (using GBM model as recommended in [52] and [53]), for clustering the K prototype function from the Python K modes library [54] was used, for machine learning models the python scikit-learn library was used, and for PCA the scikit learn's PCA function was used. The data was standardized before applying PCA. Standardization is done by subtracting the mean and scaling it to unit variance.

Synthetic data was generated from raw unprocessed real data. (as recommended by recent experiments in [44]). When generating synthetic data, default generation settings are used for all generators except in SDV and DS. SDV suffered from low efficiency, after reaching to the creators, they recommended setting the Distribution for categorical attributes as Gaussian KDE distribution to obtain more accurate results (however, this change affected its efficiency). DS suffers an efficiency problem that increased with the number of attributes, After reaching to the authors, they recommended lowering the maximum number of parents' nodes allowed, $k$, from unspecified to 3 or 2. We followed their strategy when faced with highly inefficient cases (where generation takes more than 2 days). Efficiency problems were encountered in datasets with large features even with $k = 2$, in these cases partial generation for DS was done (see Table 1).

When generating machine learning models to calculate the prediction accuracy, we preprocess synthetic and real data independently (encoding and scaling). For synthetic data, all default parameters are used and all features are included (in other words, no feature selection is performed) following recommendations in the literature [44]. Before calculating the metrics, all data are processed to handle missing values. Real and synthetic datasets are processed independently. The choice of the imputation method is explained and justified in Appendix A.

## 4 Results

We present the results into two separate subsections, the first reports on the comparison between the two utility measures $u_{PCA}$ and $p$, while the second uses the measures to evaluate the four synthetic data generators.

---

[4] In addition to prediction accuracy, we also calculated the F1 score. However we did not include the F1 scores in the analysis as they are highly correlated with accuracy values across all datasets and models.

## 4.1 Utility measures comparison

### 4.1.1 Quality effect on metric values

In this experiment, we aim to understand the effect that optimizing on the quality metrics has on the different utility dimensions. In other words, for each dataset, if we consider the optimal SD generated with respect to $u_{PCA}$ (or $p$), then how is this optimal SD performing with respect to the 3 defined utility dimensions? Moreover, which quality metric ($u_{PCA}$ or $p$) ensures better (balanced) results across all defined dimensions of utility?

To answer the above questions, we consider, for each real (testing) dataset $D_i$, all the SDs generated, and select the two optimal datasets according to each quality metric. Then we examine how well each optimal SD is performing in terms of attribute fidelity, bivariate fidelity as well as multivariate fidelity ($H, p, LC$ and $PCD$). We start first with some definitions.

**Definition 1**
We define $H_m^i, LC_m^i, p_m^i, PCD_m^i$ to be the minimal values for the four metrics across all synthetic data generated from $D_i$, in other words $X_m^i = min_{j=1,..,16}\{X_j^i\}$, for $X \in \{H, LC, p, PCD\}$.

**Definition 2**
For each quality metric $u \in \{p, u_{pca}\}$, we define a synthetic data winner on each dataset and generator and denoted by $SD_u^i(SDG_k)$. In other words, $SD_u^i(SDG_k)$ the optimal SD for the pair $\{D_i, SDG_k\}$ according to $u$.

**Definition 3**
For each datasets $D_i$, we define two (overall) optimal synthetic datasets $SD_p^i$ and $SD_{pca}^i$ as the synthetic dataset with the highest $p$ and the highest $u_{pca}$ respectively across all generators, in other words $SD_u^i = \{SD\ with\ \min u_j^i\ ; j \in \{1,\ldots,16\}; i \in\ testing\ datset\}$.

Figure 3 depicts, for each utility metric $X$, the values obtained across all 16 synthetic datasets over all testing data. The minimal values $X_m^i$ for each metric are depicted by the blue line, and the metric values corresponding to the optimal SD are depicted by the red line, for $p$ (or $SD_p^i$) [Figure 3.a], and $u_{pca}$ ($SD_{pca}^i$) [Figure 3.b]. Table 2 calculates the average of the absolute difference between $X_m^i$ and $SD_u^i$ for all $X$ and $u$.

It is evident from the graphs and the table that $u_{pca}$ generates optimal solutions with better values on all metrics apart from propensity (as expected). However, it is important to note that the average absolute difference for propensity when optimising on $u_{PCA}$ is 0.0085, i.e. within the lowest 3.5% propensity values.

| Metrics (metric range) | Average abs diff ($p$) | Average abs diff ($u_{pca}$) |
|---|---|---|
| H (0-1) | 0.0335 | 0.0052 |
| Prop (0-.025) | 0.0000 | 0.0085 |
| LC (-4.7-1.45) | 0.3847 | 0.0117 |
| PCD (0.06-85.84) | 1.1132 | 0.5587 |

Table 2. Average absolute difference between the best metric value and the corresponding metric value of the optimal models $SD_p^i$ and $SD_{pca}^i$.

In conclusion, SDs that are optimized with respect to $u_{PCA}$ provide better overall results with respect to the different utility dimensions than those optimized with respect to $p$. When accounting for the differences in metric ranges, the overall average absolute difference when optimizing with $u_{PCA}$ is 0.146025, while it is 0.38285 when optimizing with $p$.

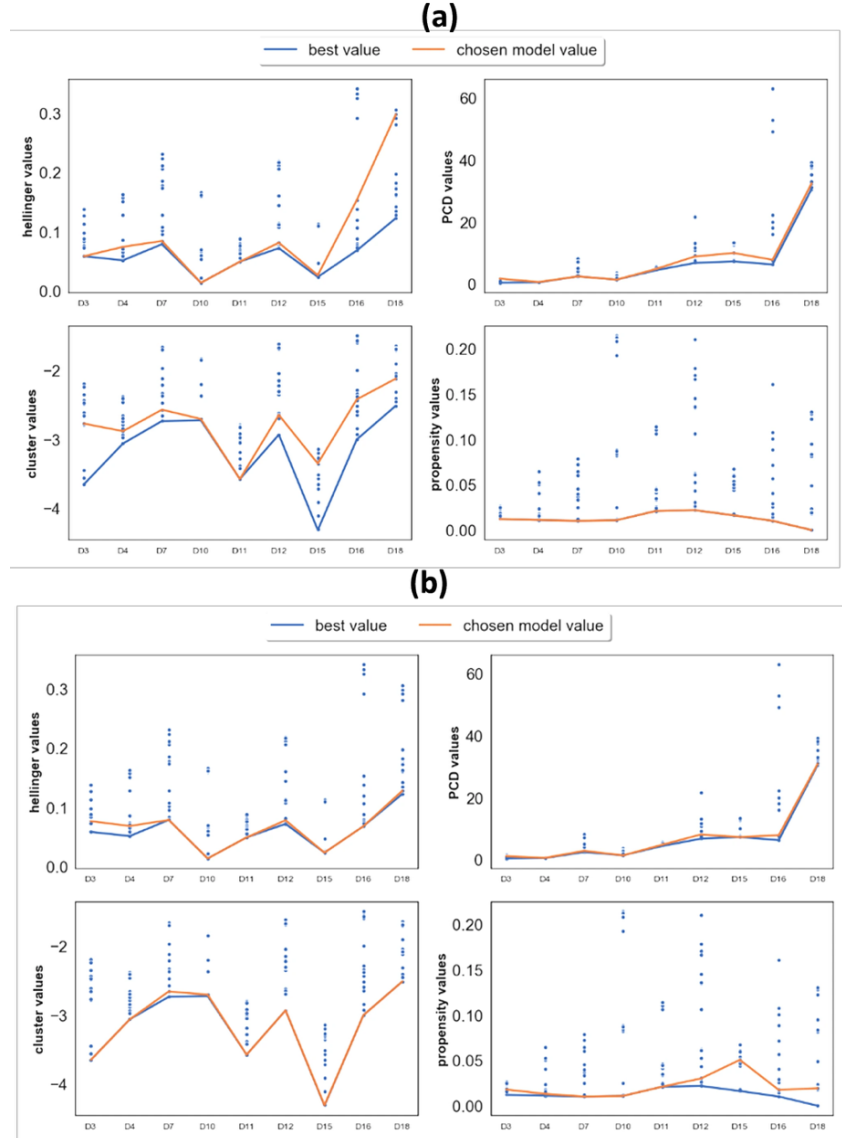

Figure 3. The graphs display the values of the different metrics for each generated SD, the lowest metric value ($X_m^i$) is displayed in blue, and the value corresponding to the optimal model is shown in red. $SD_p^i$ are displayed in (a) and $SD_{pca}^i$ in (b).

### 4.1.2 Correlation between quality and relative accuracy

Next, we consider the performance of machine learning models when applied to the synthetic datasets. In particular, we look at two measures:

1. The correlation between the quality measures and the average relative accuracy of all generated synthetic datasets across the four machine learning models. Correlation is useful in expressing the strength and direction of the relationship between the quality measures and the real-life performance of synthetic data. Correlations should be positive and value greater than 0.5 to be considered significant [55] .
2. Synthetic datasets are currently used for exploratory analysis, and it is often the case that the final analysis is performed on real data. For synthetic data to be useful under this scenario, it is important for the models generated from synthetic data to be applicable on real data. As such, a match between the winning classifiers when trained on real data and the winning classifiers when trained on synthetic data is an important measure of performance.

To compute correlations, we first define four relative accuracy measures (RA) and the average relative accuracy.

> **Definition 4**
> Given a synthetic dataset $SD_j^i$ and a CM $X$, $RA_{X,j}^i$ is defined as follows:
> $$RA_{X,j}^i = |PA_{X,j}^i - PA_X^i| \text{ where } X \in \{LR, DT, SVM, RF\}.$$
> We also define the average relative accuracy (ARA), of a synthetic data $SD_j^i$, denoted by $ARA_j^i$ as the average of the four RA measures as follows:
> $$ARA_j^i = \frac{1}{4}\sum_X RA_{X,j}^i \text{ where } X \in \{LR, DT, SVM, RF\}.$$
> Thus, each synthetic data is associated with an ARA.

The correlations between the quality metrics and ARA relative to every synthetic data generator are displayed in Table 3. It is evident that correlations with $u_{PCA}$ is stronger while maintaining the same direction (positive) across all generators.

|       | $p$    | $u_{pca}$ |
|-------|--------|-----------|
| DS    | 0.046  | 0.537     |
| SDV   | -0.308 | 0.548     |
| SP-np | 0.575  | 0.670     |
| SP-p  | 0.663  | 0.708     |

Table 3. Correlation between quality metrics and average relative accuracy

Table 4 displays the overall correlation between quality metrics and machine learning performance. In other words, for each synthetic dataset, we consider the quality and performance metrics associated with it regardless of how it was generated. This overall correlation gives us an idea about the association between quality and performance of any masked data irrespective of how the data was generated. The p-values for the correlations are also presented in the table, they indicate high statistical significance for $u_{pca}$'s correlation with performance and no significant correlation between performance and $p$.

|             | $p$   | $u_{pca}$ |
|-------------|-------|-----------|
| Correlation | 0.006 | 0.525     |
| p-value     | 0.947 | 1.38e-11  |

Table 4. Overall correlation between quality metrics and average relative accuracy along with p-values

Evidently, $u_{pca}$ presents significantly higher correlations than $p$ across all generators. The strength of correlation is moderate to strong. The overall correlation across all generated synthetic dataset (irrespective of generator) is moderate for $pca$ and insignificant for $p$.

4.1.3 Matches on winning classfier

Denote the winning classifier for $SD_p^i$ and $SD_{pca}^i$ by $WC_p^i$ and $WC_{pca}^i$ respectively (i.e. for both optimal datasets, we consider the classifier that presented the highest model accuracy in the 9 testing datasets). Table 5 counts the number of times each of $WC_p^i$ and $WC_{pca}^i$ matches the winning classifier on the corresponding real dataset $D_i$. As displayed, $u_{pca}$ lead to 22% more matches with the real on the winning classifier.

| $p$ | $u_{pca}$ |
|---|---|
| 5/9=55.56% | 7/9=77.78% |

Table 5. Matches on winning classifier

## 4.2 Best Synthesizer

In this section we compare the four synthetic data generators on multiple aspects: (i) stability, (ii) supervised machine learning accuracy, and (iii) winning classifier matches with real.

### 4.2.1 Stability

The stability of a synthetic data generator reports on the consistency of the estimates across the synthetic datasets generated from the same real dataset. As in [5], the stability (with respect to both quality measures) is calculated as the range of quality values across all SDs generated from the same real data for a given generator. The overall stability of a generator is then calculated as the average of all stabilities across the different datasets (see Definition 5).

**Definition 5**

For $u \in \{p, u_{pca}\}$, the stability of $SDG_k$ on datasets generated from $D_i$, denoted by $S_u^i(SDG_k)$, is defined as the range of $u$ values:
$$S_u^i(SDG_k) = range_{j\ corresponding\ to\ SDG_k}[u\ of\ SD_j^i].$$
The overall stability for $SDG_k$ is defined as the average across all datasets:
$$S_u(SDG_k) = average_i S_u^i(SDG_k).$$

Table 6 presents the overall stability for each generator for both quality measures.

| SDGs | $p$ | $u_{pca}$ |
|---|---|---|
| DS | 0.0183 | 0.3982 |
| SDV | 0.0297 | 0.3098 |
| SP-np | **0.0059** | 0.3653 |
| SP-p | 0.0183 | **0.2821** |

Table 6. Average stability

In addition to the above, for each dataset $D_i$ and synthetic data generator $SDG_k$, we calculate the width of the confidence interval for each ensemble using 95% t-distribution test. The average of the (four) widths is then associated with dataset $D_i$ and $SDG_k$ for $p$ and $u_{pca}$ respectively. The values are shown in Figure 4.

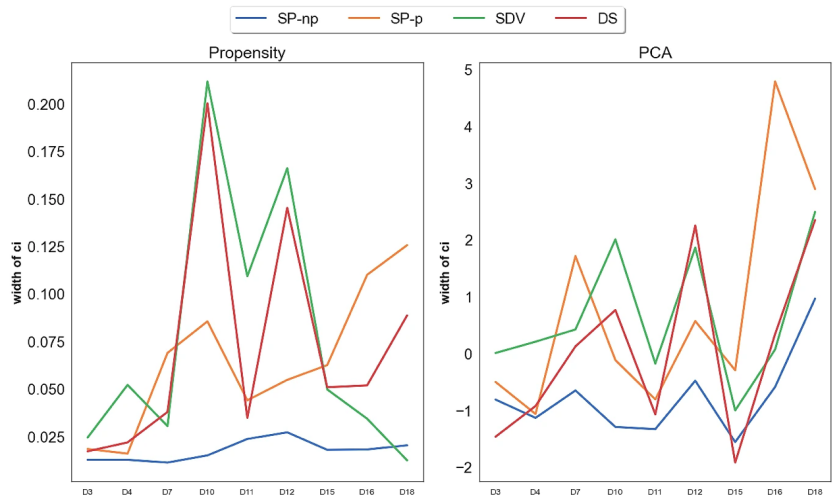

Figure 4. Average of confidence interval widths for both quality measures

The average stability and average confidence interval widths provide us with information regarding consistency in the utility of the synthetic data produced. From Table 4, It is evident that SP-np has the best average stability on the $p$ metric with an average range value of 2.36% of the overall range of $p$ (while the least stability is reported by SDV and is about 12% of the overall range values. For the $pca$ metric, the generators have close stability values; they are all within 1% of the overall $u_{pca}$ range (the overall $u_{pca}$ range is [-2.55, 44.48]). As for Figure 4, it is evident that SP-np shows the least variation in confidence interval widths for both quality measures.

4.2.2 ML accuracy and winning synthesizer

In this subsection, we aim to identify the generator that is associated with the best performance. To achive this, we perform the following three analyses:
1. We consider, for each generator, the highest quality synthetic datasets (per $p$, and $u_{pca}$), then we compare their accuracy.
2. For each dataset, we identify the generator that produced the dataset with the highest quality (per $p$, and $u_{pca}$).
3. For each generator, we calculate the number of winning classifiers matches with the corresponding real data. As mentioned earlier, this analysis is important as we need the models generated from synthetic data to be applicable on real data. The analysis in Section 4.1.3 looked at the overall matches irrespective of the generators, while in this section we compare the matches per generator.

For the first analysis, we consider the relative accuracy values for the optimal synthetic data $SD_p^i$ and $SD_{pca}^i$; $i \in testing$, denoted by $RA_{X,p}^i$ and $RA_{X,pca}^i$; $X \in \{LR, DT, SVM, RF\}$ respectively. Figure 5 depicts the average of these RA values across all testing datasets $Average_i(RA_{X,u}^i); u \in \{p, u_{pca}\}$. This measure helps determine the SDG that provides the best real-life performance in two cases: 1) when propensity is used to select the best SD generated and 2) when $pca$ is used.

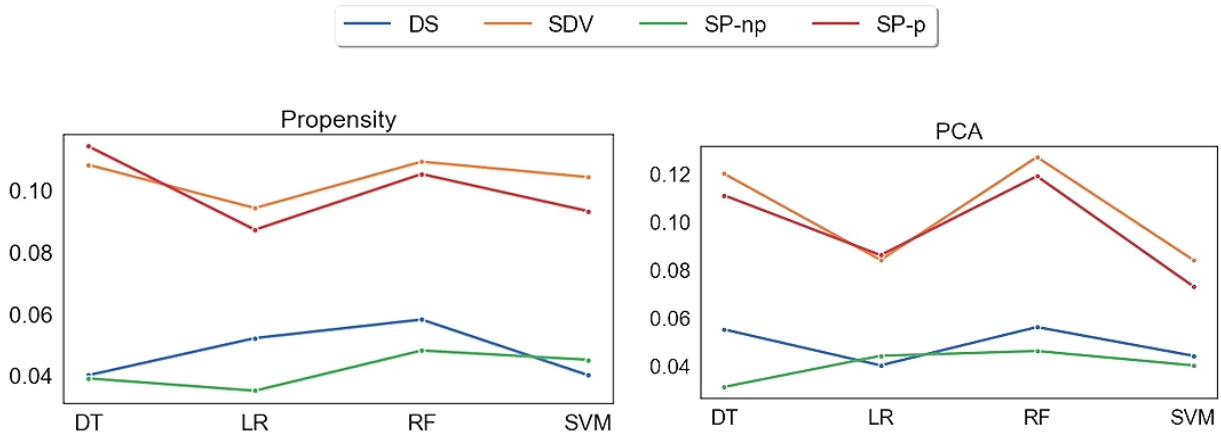

Figure 5. RA for the winning synthesizers

It is evident that $SP - np$ produces the best SD for both quality measures, followed by $DS$.

Next, Table 7 shows the number of times each SDG generated the optimal SD according to $p$ and $u_{pca}$ respectively. These results help identifying the generator that produces datasets with the highest quality. $SP - np$ produced the best SD 77.79% of the time according to both measures.

|       | $p$     | $u_{pca}$ |
|-------|---------|-----------|
| DS    | 0%      | 22.21%    |
| SDV   | 22.21%  | 0%        |
| SP-np | 77.79%  | 77.79%    |
| SP-p | 0% | 0% |

Table 7. Winning SD generators

Finally, Table 8 displays the number of times the winning classifiers for the SDs matched that of the corresponding real datasets across all generators. The table shows a better performance for SP-np followed by DS.

| SDGs | Matches on winning classifier |
|---|---|
| DS | 21/36=58.33 % |
| SDV | 14/36=38.89 % |
| SP-np | 23/36=63.89 % |
| SP-p | 12/36=33.33 % |

Table 8. Matches on winning classifiers between SDs and the corresponding real datasets

# 5 Conclusion

Synthetic data is rapidly emerging as a valuable technology for sharing data for secondary purposes. It was originally proposed by Rubin in 1993 [56], but has recently gained traction due to increased calls for broader, faster and more democratic data sharing. Moreover, recent advances in machine learning led to the introduction of newer methods believed to construct better quality synthetic data [57].

Multiple broad utility measures have been defined for assessing the overall quality of synthetic data. These measures have been found to generate conflicting conclusions making direct comparison of synthetic data generators surprisingly difficult. In fact, prior research found weak or no correlation between some of the most popular (broad) metrics available, concluding that these measures concentrate on different utility dimensions and none can be used to assess the overall utility of synthetic data.

This paper aggregates four (non-correlated) utility metrics into one measure of utility using PCA and checks whether the generated measure can be used to generate SDs that perform well in real life in comparison with Propensity, a commonly used measure of overall SD quality.

The results suggest that datasets with higher $u_{pca}$ fare better on univariate, bivariate and multivariate fidelity, and are better correlated with (supervised) machine learning accuracy. Moreover, when synthetic data generation is optimized by $u_{pca}$, the classifier that performs best on the SD matches that of the real data in about 77% of the cases (versus 55% for datasets with optimal propensity)

The paper also uses the newly generated measure to compare four well-recognized synthetic data generators, two of which are based in the statistical domain and two in machine learning. The results suggest that SP-np generates the highest quality SD about 78% of the time (followed by DS 22% of the time), and produces machine learning models with the highest relative accuracy (followed by DS). Moreover, SP-np matches the winning classifiers of the real about 64% of the time (followed by DS 58%).

In terms of limitations, further investigations with more datasets, more datapoints (per ensemble), and additional machine learning algorithms are needed to:

- Validate the above results
- Make sure that the results are independent of the particular machine learning algorithms used and
- To refine the eigenvectors for the PCA based measure.

Further investigations into the best broad measures to include in our quality measure are also needed to strengthen the conclusions.

# Appendix A

Before calculating any metric, all data are processed to handle missing values. Real and synthetic datasets are processed independently as shown in the figure below.

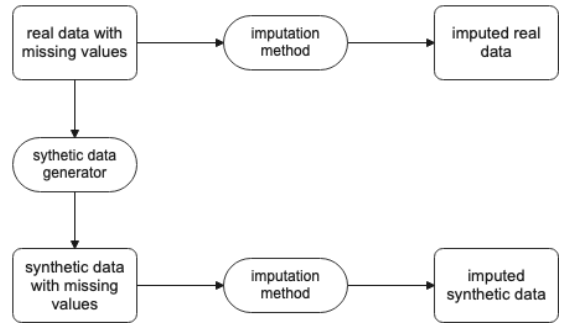

Figure B1. Independent imputation

In order to choose an appropriate imputation method, we performed a comparison between 3 commonly referenced single imputation methods:

1. Regression imputation method (RI): this method fits a regression model of the variable containing the missing values with other variables as predictors. The obtained model is used to impute the missing values, and then some (calculated) random noise is added to that value [58].

2. Cart imputation method (CI): this method fits a tree model to the data, with the variable containing missing values as the target variable. The terminal node for the missing value is then determined by fitting the tree with the associated attribute values, where a value is chosen. Finally, a parameter of uncertainty is incorporated into the chosen value [59].

3. Mean and mode Substitution method (SI): This is one of the simplest methods, it involves filling the missing value with the mean (in case of continuous attributes) or mode (in case of categorical attributes) of the column to which the missing data belongs [60].

We apply the 3 imputation methods to 4 datasets (dermatology, diabetes, mice and titanic from Table 1 in the main text) and compared the differences in ranges between the real and the synthetic data.

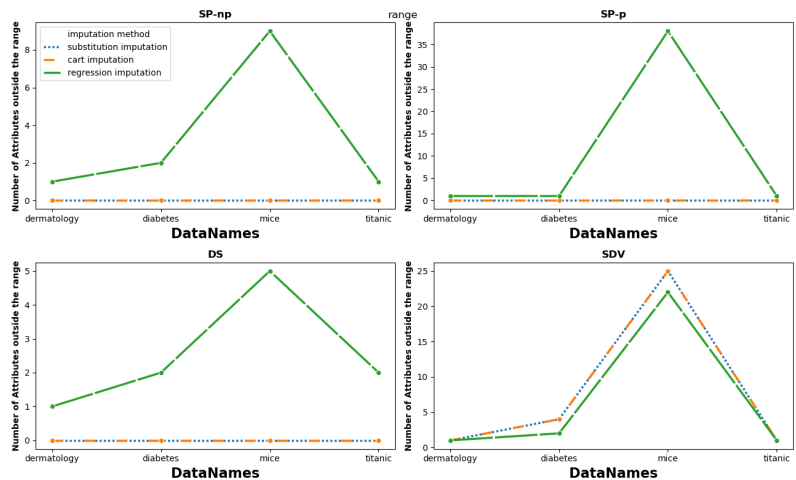

Figure B2. Number of attributes in synthetic data that are outside the range of the real data

Figure B2 displays the results. For SP-np, SP-p, and DS, the synthetic datasets contain attributes with values outside the real when the regression imputation method is used. For the other two methods, there are no attributes outside the range. Figure B2 clearly shows that RI introduces outliers to the imputed data that are not present in the real data which led to its exclusion.

To compare SI and CI, we use two utility metrics, Hellinger and Propensity. The first 9 datasets ($D_1, \ldots, D_9$) from Table 1 were used for this comparison. The results are depicted in Figures B3 (for hellinger) and B4 (for propensity).

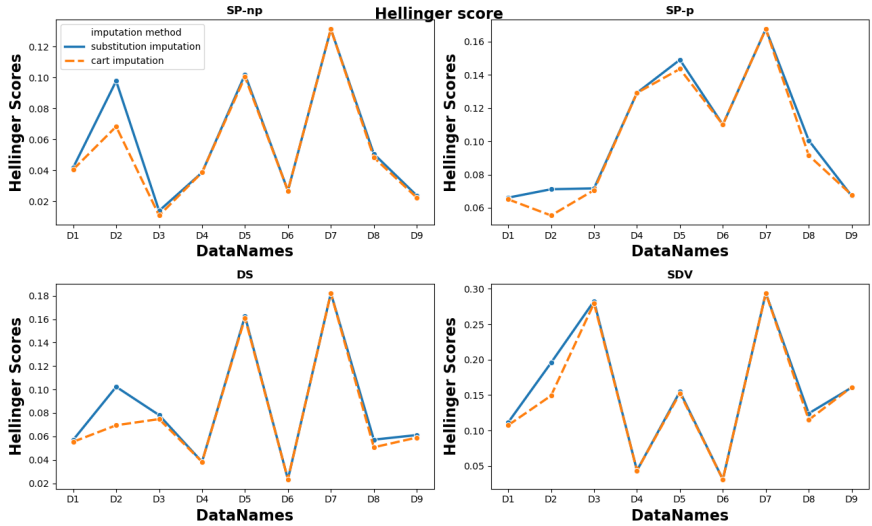

Figure B3. Hellinger Scores

As shown if Figure B3, Hellinger scores are very close for the two methods, there are some cases where the cart imputation performs better (by displaying lower Hellinger score for D2 and D8).

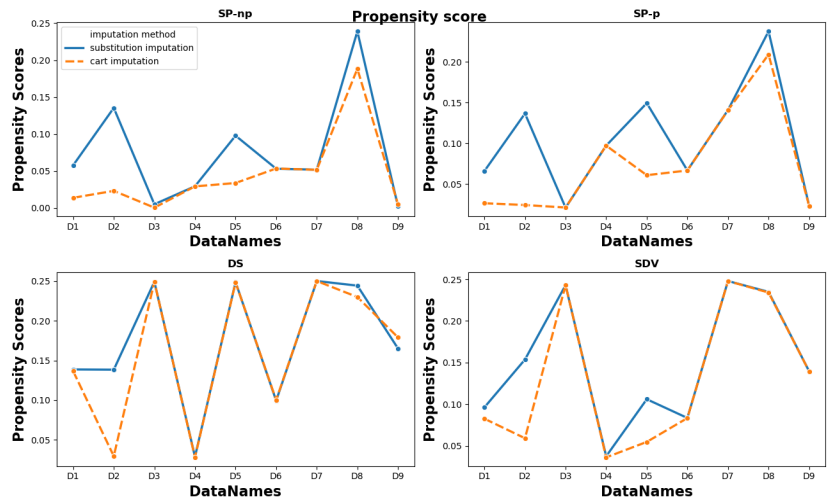

Figure B4. Propensity Scores

Similarly, the two imputation methods behave closely in terms of propensity score. However, there are multiple datasets when the cart imputation outperforms the substitution method. The results from both experiments justify the usage of the cart imputation over the other two.